\documentclass[twocolumn,english,aps,prl,showpacs]{revtex4}
\usepackage[T1]{fontenc}
\usepackage[latin1]{inputenc}
\usepackage{amsmath}
\usepackage{graphicx}
\usepackage{amssymb}
\usepackage{esint}

\makeatletter
\@ifundefined{textcolor}{}
{%
 \definecolor{BLACK}{gray}{0}
 \definecolor{WHITE}{gray}{1}
 \definecolor{RED}{rgb}{1,0,0}
 \definecolor{GREEN}{rgb}{0,1,0}
 \definecolor{BLUE}{rgb}{0,0,1}
 \definecolor{CYAN}{cmyk}{1,0,0,0}
 \definecolor{MAGENTA}{cmyk}{0,1,0,0}
 \definecolor{YELLOW}{cmyk}{0,0,1,0}
 }

\makeatother

\makeatother

\usepackage{babel}

\makeatother

\usepackage{babel}

\makeatother

\usepackage{babel}

\begin{document}

\title{Dynamic response of strongly correlated Fermi gases in the quantum
virial expansion}

\author{Hui Hu$^{1,2}$,}

\thanks{To whom correspondence should be addressed. E-mail: hhu@swin.edu.au}

\author{Xia-Ji Liu$^{1}$, and Peter D. Drummond$^{1}$}

\affiliation{$^{1}$\ ARC Centre of Excellence for Quantum-Atom Optics, Centre
for Atom Optics and \\
 Ultrafast Spectroscopy, Swinburne University of Technology, Melbourne
3122, Australia, \\
 $^{2}$\ Department of Physics, Renmin University of China, Beijing
100872, China}

\date{\today{}}
\begin{abstract}
By developing a quantum virial expansion theory, we quantitatively
calculate the dynamic density response function of a trapped strongly
interacting Fermi gas at high temperatures near unitarity. A clear
transition from atomic to molecular responses is identified in the
spectra when crossing from the BCS to BEC regimes, in qualitative
agreement with recent Bragg spectroscopy observations. Our virial
expansion method provides a promising way to solve the challenging
strong-coupling problems and is applicable to other dynamical properties
of strongly correlated Fermi gases. 
\end{abstract}


\maketitle

\section{Introduction}

The dynamic structure factor (DSF) plays a fundamental role in understanding
quantum many-body systems \cite{allanbook}: it gives the response
of the system to an excitation process that couples to density. Experimental
advances in Bragg spectroscopy have now measured the DSF for strongly
interacting ultra-cold fermions. At a high momentum transfer, these
experiments show evidence for a clear transition from atomic to molecular
response when traversing from the BCS to BEC regimes\cite{swinexpt}.
However, while the DSF for weakly interacting bosons is well-studied
theoretically and experimentally \cite{braggbec}, the expected dynamic
structure factor of strongly interacting fermions remains unknown,
except for a dynamical mean-field approximation \cite{combescot,minguzzi}.
This is due to the notorious absence of a small parameter for strongly
interacting particles. The Bragg experiments create a theoretical
challenge: how can one reliably calculate the dynamic structure factor
in this strongly interacting regime? Similar challenges also arise
when one tries to understand some recent measurements on radiofrequency
spectroscopy of strong interacting fermions \cite{rfgrimm,rfketterle,rfzwierlein}.

In this paper, we present a systematic study of the dynamic structure
factor of a\emph{ normal}, \emph{trapped} and \emph{strongly interacting}
Fermi gas at high temperatures. This is achieved by developing a quantum
virial expansion for dynamical properties of many-body systems. Our
expansion is applicable to arbitrary interaction strengths and has
a controllable small parameter. The fugacity $z\equiv\exp(\mu/k_{B}T)\ll1$
is small since the chemical potential $\mu$ tends to $-\infty$ at
large temperatures $T$. We note that dynamical studies using virial
expansions have been carried out for classical, untrapped gases \cite{Miyazaki}.
However, previous virial expansion studies of quantum gases were restricted
to static properties only \cite{jasonho,ourve}. By comparing the
virial expansion prediction \cite{unitaritycmp} with the experimentally
measured equation of state \cite{nascimbene}, we find a wide applicability
of the expansion for a trapped Fermi gas: it is valid down to temperatures
as low as 0.4$T_{F}$ (see Ref. \cite{unitaritycmp} for details).
Here, $T_{F}$ is the Fermi temperature of a trapped ideal, non-interacting
Fermi gas.

Our main results may be summarized as follows. We find a smooth transition
in the dynamic structure factor, from an atomic response to a molecular
response as the interaction strength increases (Fig. 1). This feature
agrees reasonably well with recent experimental measurements although
the latter was carried out at lower temperatures. We show that the
spin-antiparallel dynamic structure factor provides the most sensitive
probe for molecule formation (Figs. 2 and 3). The static structure
factor is also obtained as a subset of our results (Fig. 4). These
predictions are readily testable experimentally.

\section{Quantum virial expansion of dynamic structure factor}

We start by constructing the virial expansion for the dynamic structure
factor $S({\bf q},\omega)$, the Fourier transform of the density-density
correlation functions at two different space-time points. Consider
a harmonically trapped atomic Fermi gas with an equal number of atoms
($N/2$) in two hyperfine states (referred to as spin-up, $\sigma=\uparrow$,
and spin-down, $\sigma=\downarrow$), where $S_{\uparrow\uparrow}({\bf q},\omega)=S_{\downarrow\downarrow}({\bf q},\omega)$
and $S_{\uparrow\downarrow}({\bf q},\omega)=S_{\downarrow\uparrow}({\bf q},\omega)$.
To calculate the total dynamic structure factor $S({\bf q},\omega)\equiv2[S_{\uparrow\uparrow}({\bf q},\omega)+S_{\uparrow\downarrow}({\bf q},\omega)]$,
it is convenient to work with the dynamic susceptibility \cite{allanbook},
$\chi_{\sigma\sigma^{\prime}}\left({\bf r},{\bf r}^{\prime};\tau\right)\equiv-\left\langle T_{\tau}\hat{n}_{\sigma}\left({\bf r},\tau\right)\hat{n}_{\sigma^{\prime}}\left({\bf r}^{\prime},0\right)\right\rangle $,
where $\hat{n}_{\sigma}\left({\bf r},\tau\right)$ is the density
(fluctuation) operator in spin channel $\sigma$, and $\tau$ is an
imaginary time in the interval $0<\tau\leq\beta=1/k_{B}T$.

The DSF is then obtained \cite{allanbook} from the Fourier components
$\chi_{\sigma\sigma^{\prime}}\left({\bf r},{\bf r}^{\prime};i\omega_{n}\right)$
at discrete Matsubara imaginary frequencies $i\omega_{n}=i2n\pi k_{B}T$
($n=0,\pm1,...$), via analytic continuation and the fluctuation-dissipation
theorem: \[
S_{\sigma\sigma^{\prime}}\left({\bf r},{\bf r}^{\prime};\omega\right)=-\frac{\mathop{\rm Im}\chi_{\sigma\sigma^{\prime}}\left({\bf r},{\bf r}^{\prime};i\omega_{n}\rightarrow\omega+i0^{+}\right)}{\pi(1-e^{-\beta\omega})}\,\,\,.\]
 A final Fourier transform with respect to the relative spatial coordinate
${\bf r}-{\bf r}^{\prime}$ leads to $S_{\sigma\sigma^{\prime}}\left({\bf q},\omega\right)$.

Our quantum virial expansion applies to the dynamic susceptibility
$\chi_{\sigma\sigma^{\prime}}\left({\bf r},{\bf r}^{\prime};\tau>0\right)$,
which is formally expanded as: \begin{equation}
\chi_{\sigma\sigma^{\prime}}\equiv-\frac{\text{Tr}\left[e^{-\beta\left({\cal H}-\mu{\cal N}\right)}e^{{\cal H}\tau}\hat{n}_{\sigma}\left({\bf r}\right)e^{-{\cal H}\tau}\hat{n}_{\sigma^{\prime}}\left({\bf r}^{\prime}\right)\right]}{\text{Tr}e^{-\beta\left({\cal H}-\mu{\cal N}\right)}}.\end{equation}
 At high temperatures, Taylor-expanding in terms of the powers of
small fugacity $z\equiv\exp(\mu/k_{B}T)\ll1$ leads to $\chi_{\sigma\sigma^{\prime}}\left({\bf r},{\bf r}^{\prime};\tau\right)=(zX_{1}+z^{2}X_{2}+\cdots)/(1+zQ_{1}+z^{2}Q_{2}+\cdots)=zX_{1}+z^{2}\left(X_{2}-X_{1}Q_{1}\right)+\cdots$,
where we have introduced the cluster functions $X_{n}=-$ Tr$_{n}[e^{-\beta{\cal H}}e^{\tau{\cal H}}\hat{n}_{\sigma}({\bf r)}e^{-\tau{\cal H}}\hat{n}_{\sigma^{\prime}}({\bf r}^{\prime})]$
and $Q_{n}=$Tr$_{n}[e^{-\beta{\cal H}}]$, with $n$ denoting the
number of particles in the cluster and Tr$_{n}$ denoting the trace
over $n$-particle states of proper symmetry. We shall refer to the
above expansion as the virial expansion of dynamic susceptibilities,
$\chi_{\sigma\sigma^{\prime}}\left({\bf r},{\bf r}^{\prime};\tau\right)=z\chi_{\sigma\sigma^{\prime},1}\left({\bf r},{\bf r}^{\prime};\tau\right)+z^{2}\chi_{\sigma\sigma^{\prime},2}\left({\bf r},{\bf r}^{\prime};\tau\right)+\cdots,$
where, \begin{eqnarray}
\chi_{\sigma\sigma^{\prime},1}\left({\bf r},{\bf r}^{\prime};\tau\right) & = & X_{1},\nonumber \\
\chi_{\sigma\sigma^{\prime},2}\left({\bf r},{\bf r}^{\prime};\tau\right) & = & X_{2}-X_{1}Q_{1},\ \text{etc}.\end{eqnarray}
 Accordingly, we shall write for the dynamic structure factors, $S_{\sigma\sigma^{\prime}}\left({\bf q},\omega\right)=zS_{\sigma\sigma^{\prime},1}\left({\bf q},\omega\right)+z^{2}S_{\sigma\sigma^{\prime},2}\left({\bf q},\omega\right)+\cdots$.
It is readily seen that a similar virial expansion holds for other
dynamical properties. As anticipated, the determination of the $n$-th
expansion coefficient requires the knowledge of all solutions up to
$n$-body, including both the eigenvalues and eigenstates. Here we
aim to calculate the leading effect of interactions, which contribute
to the 2nd-order expansion function. For this purpose, it is convenient
to define $\Delta\chi_{\sigma\sigma^{\prime},2}\equiv\left\{ \chi_{\sigma\sigma^{\prime},2}\right\} ^{(I)}=\left\{ X_{2}\right\} ^{(I)}$
and $\Delta S_{\sigma\sigma^{\prime},2}\equiv\left\{ S_{\sigma\sigma^{\prime},2}\right\} ^{(I)}$.
The notation $\left\{ \right\} ^{(I)}$ means the contribution due
to interactions inside the bracketed term, so that $\left\{ X_{2}\right\} ^{(I)}=X_{2}-X_{2}^{(1)}$,
where the superscript {}``1'' in $X_{2}^{(1)}$ denotes quantities
for a noninteracting system. We note that the inclusion of the 3rd-order
expansion function is straightforward, though involving more numerical
effort.

To solve the two-fermion problem, we adopt a short-range \textit{S}-wave
pseudopotential for interactions between two fermions with {\em
opposite} spins, in accord with the experimental situation of broad
Feshbach resonances. In an isotropic harmonic trap with potential
$V\left({\bf r}\right)=m\omega_{0}^{2}r^{2}/2$, the solution is known
\cite{ourve}. Any eigenstate with energy $E_{P}=\epsilon_{p1}+\epsilon_{p2}$
($P\equiv\{p1,p2\}$) can be separated into center-of-mass and relative
motions, $\Phi_{P}({\bf \textbf{r}},\mathbf{r}')=\varphi_{p1}([{\bf \textbf{r}}+\mathbf{r}']/2)\psi_{p2}({\bf \textbf{r}}-\mathbf{r}')$.
The center-of-mass wave function is not affected by interactions,
according to Kohn's theorem: it is simply the single-particle wave
function of a three-dimensional isotropic harmonic oscillator, but
with mass $M=2m$.

For the \emph{relative} wave function with a quantum number $p2\equiv\{n_{p},l_{p}\}$,
only the branch with zero relative angular momentum ($l_{p}=0$) is
modified by interactions. The relative energy $\epsilon_{p2}\equiv(2\nu_{n_{p}}+3/2)\hbar\omega_{0}$
is determined by $\sqrt{2}\Gamma(-\nu_{n_{p}})/\Gamma(-\nu_{n_{p}}-1/2)=a_{ho}/a$,
where $a_{ho}\equiv\sqrt{\hbar/m\omega_{0}}$ is the characteristic
oscillator length of the external trap potential, and $a$ is the
\textit{S}-wave scattering length. The relative wave function is then
given by $\psi_{p2}({\bf x})=A_{n_{p}}\exp(-x^{2}/4a_{ho}^{2})\Gamma(-\nu_{n_{p}})U(-\nu_{n_{p}},3/2;x^{2}/2a_{ho}^{2})$,
with $A_{n_{p}}$ being the normalization factor. Here, $\Gamma$
and $U$ are the Gamma function and confluent hypergeometric function,
respectively. Other branches with $l_{p}\neq0$ (together with the
non-interacting counterpart $\psi_{p2}^{\left(1\right)}({\bf x})$
for all $l_{p}$) are given by the standard single-particle wave function
of a harmonic oscillator, with a reduced mass $m/2$.

With this backgrounds, we turn to consider the 2nd-order expansion
function for the dynamic susceptibility, $\Delta\chi_{\sigma\sigma^{\prime},2}=-\left\{ Tr_{\uparrow\downarrow}\left[e^{-\beta{\cal H}}e^{\tau{\cal H}}\hat{n}_{\sigma}\left({\bf r}\right)e^{-\tau{\cal H}}\hat{n}_{\sigma^{\prime}}\left({\bf r}^{\prime}\right)\right]\right\} ^{(I)}$.

The trace is calculated by inserting the identity $\sum_{Q}\left|Q\right\rangle \left\langle Q\right|={\bf \hat{1}}$.
We find $\Delta\chi_{\sigma\sigma^{\prime},2}=-\sum_{P,Q}\left\{ e^{-\beta E_{P}+\tau(E_{P}-E_{Q})}\left\langle P\left|\hat{n}_{\sigma}\right|Q\right\rangle \left\langle Q\left|\hat{n}_{\sigma^{\prime}}\right|P\right\rangle \right\} ^{(I)}$.
The sum is over all the pair states $P$ and $Q$ with energies $E_{P}$
and $E_{Q}$. Expressing the density operator in first quantization:
$\hat{n}_{\uparrow}\left({\bf r}\right)=\sum_{i}\delta\left({\bf r}-{\bf r}_{i\uparrow}\right)$
and $\hat{n}_{\downarrow}\left({\bf r}\right)=\sum_{j}\delta({\bf r}-{\bf r}_{j\downarrow})$,
it is straightforward to show that, \begin{equation}
\Delta\chi_{\sigma\sigma^{\prime},2}=-\sum_{P,Q}\left\{ e^{-\beta E_{P}+\tau\left(E_{P}-E_{Q}\right)}C_{\sigma\sigma^{\prime}}^{PQ}\left({\bf r},{\bf r}^{\prime}\right)\right\} ^{(I)},\end{equation}
 where $C_{\uparrow\uparrow}^{PQ}\equiv\int d{\bf r}_{2}d{\bf r}_{2}^{\prime}\left[\Phi_{P}^{*}\Phi_{Q}\right]\left({\bf r},{\bf r}_{2}\right)[\Phi_{Q}^{*}\Phi_{P}]\left({\bf r}^{\prime},{\bf r}_{2}^{\prime}\right)$
and $C_{\uparrow\downarrow}^{PQ}\equiv\int d{\bf r}_{1}d{\bf r}_{2}[\Phi_{P}^{*}\Phi_{Q}]\left({\bf r},{\bf r}_{2}\right)[\Phi_{Q}^{*}\Phi_{P}]\left({\bf r}_{1},{\bf r}^{\prime}\right)$.
The dynamic structure factor can be obtained by the analytic continuation,
giving the result that $\Delta S_{\sigma\sigma^{\prime},2}\left({\bf r},{\bf r}^{\prime};\omega\right)=\sum_{P,Q}\left\{ \delta\left(\omega+E_{P}-E_{Q}\right)e^{-\beta E_{P}}C_{\sigma\sigma^{\prime}}^{PQ}\left({\bf r},{\bf r}^{\prime}\right)\right\} ^{(I)}$.
Applying a further Fourier transform with respect to ${\bf x}$ and
integrating over ${\bf R}$, we obtain the response $\Delta S_{\sigma\sigma^{\prime},2}\left({\bf q},\omega\right)$,
\begin{equation}
\Delta S_{\sigma\sigma^{\prime},2}=\sum_{P,Q}\left\{ \delta\left(\omega+E_{P}-E_{Q}\right)e^{-\beta E_{P}}F_{\sigma\sigma^{\prime}}^{PQ}\left({\bf q}\right)\right\} ^{(I)},\end{equation}
 where $F_{\sigma\sigma^{\prime}}^{PQ}\left({\bf q}\right)=\int d{\bf r}d{\bf r}^{\prime}e^{-i{\bf q\cdot}({\bf r}-{\bf r}^{\prime})}C_{\sigma\sigma^{\prime}}^{PQ}\left({\bf r},{\bf r}^{\prime}\right)$.

To proceed, one notices that $F_{\sigma\sigma^{\prime}}^{PQ}\left({\bf q}\right)$
can be separated into center-of-mass and relative motion parts. We
thus introduce $\delta\left(\omega+E_{P}-E_{Q}\right)e^{-\beta E_{P}}\equiv\int d\omega^{\prime}\delta(\omega^{\prime}+\epsilon_{p1}-\epsilon_{q1})\delta(\omega-\omega^{\prime}+\epsilon_{p2}-\epsilon_{q2})e^{-\beta(\epsilon_{p1}+\epsilon_{p2})}$
to rewrite $\Delta S_{\sigma\sigma^{\prime},2}=\left\{ \int d\omega^{\prime}W_{cm}({\bf q},\omega^{\prime})W_{rel}^{\sigma\sigma^{\prime}}({\bf q},\omega-\omega^{\prime})\right\} ^{(I)}$,
where $W_{cm}\equiv\sum_{p1q1}\delta(\omega^{\prime}+\epsilon_{p1}-\epsilon_{q1})e^{-\beta\epsilon_{p1}}\left|f_{p1q1}\right|^{2}$
and $W_{rel}^{\sigma\sigma^{\prime}}\equiv\sum_{p2q2}\delta(\omega-\omega^{\prime}+\epsilon_{p2}-\epsilon_{q2})e^{-\beta\epsilon_{p2}}A_{p2q2}^{\sigma\sigma^{\prime}}$,
with $f_{p1q1}\equiv\int d{\bf R}e^{-i{\bf q}\cdot{\bf R}}\varphi_{p1}^{*}({\bf R)}\varphi_{q1}({\bf R)}$,
$A_{p2q2}^{\uparrow\uparrow}\equiv\left|A_{p2q2}\right|^{2}$ and
$A_{p2q2}^{\uparrow\downarrow}\equiv A_{p2q2}^{2}$, and $A_{p2q2}\equiv\int d{\bf x}e^{-i{\bf q}\cdot{\bf x}/2}\psi_{p2}^{*}({\bf x)}\psi_{q2}({\bf x)}$.
At high temperatures, we may apply a semi-classical Thomas-Fermi approximation
for the calculation of $W_{cm}$. In contrast, $W_{rel}^{\sigma\sigma^{\prime}}$
has to be summed over all the eigenstates since the relative wave
functions could be spatially singular due to strong interactions.

After some algebra, we find $W_{cm}=B\sqrt{m/\pi}\exp[-\beta(\omega^{\prime}-\omega_{R}/2)^{2}/(2\omega_{R})]$,
with a constant $B\equiv(k_{B}T)^{5/2}/(q\hbar^{4}\omega_{0}^{3})$
and $\omega_{R}\equiv\hbar q^{2}/(2m)$ being the recoil frequency
for atoms, and \begin{eqnarray}
A_{p2q2}^{\sigma\sigma^{\prime}} & = & (-1)^{l(1-\delta_{\sigma\sigma^{\prime}})}(2l+1)\times\nonumber \\
 &  & \left[\int_{0}^{\infty}dxx^{2}j_{l}\left(\frac{qx}{2}\right)\phi_{n_{p}l_{p}}\left(x\right)\phi_{n_{q}l_{q}}\left(x\right)\right]^{2},\label{Arel}\end{eqnarray}
 where we specify $p2=\{n_{p}l_{p}\}$ and $q2=\{n_{q}l_{q}\}$, and
$l=\max\{l_{p},l_{q}\}$. Here, $j_{l}$ is the spherical Bessel function
and $\phi$ is the relative {\em radial} wave function that can
be obtained from the two-fermion solution. We require that either
$l_{p}$ or $l_{q}$ should be zero (i.e., $\min\{l_{p},l_{q}\}=0$),
otherwise $A_{p2q2}^{\sigma\sigma^{\prime}}$ will be cancelled exactly
by the non-interacting terms. Inserting the expression for $W_{cm}$
and Eq. (\ref{Arel}) into $\Delta S_{\sigma\sigma^{\prime},2}$,
we finally arrive at ($\tilde{\omega}=\omega-\omega_{R}/2$), \begin{equation}
\Delta S_{\sigma\sigma^{\prime},2}=B\sqrt{\frac{m}{\pi}}\sum_{p2q2}\left\{ e^{-\frac{\beta\left(\tilde{\omega}+\epsilon_{p2}-\epsilon_{q2}\right)^{2}}{2\omega_{R}}}e^{-\beta\epsilon_{p2}}A_{p2q2}^{\sigma\sigma^{\prime}}\right\} ^{(I)}\,\,.\label{ddsf2}\end{equation}
 Eq. (\ref{ddsf2}) is the main result of this work. Together with
the non-interacting structure factor at large $T$, $S_{\sigma\sigma^{\prime},1}^{\left(1\right)}=\delta_{\sigma\sigma^{\prime}}B\sqrt{m/(2\pi)}e^{-\beta(\omega-\omega_{R})^{2}/(4\omega_{R})}$
and $S_{\sigma\sigma^{\prime},2}^{\left(1\right)}=-\delta_{\sigma\sigma^{\prime}}B\sqrt{m/(16\pi)}e^{-\beta(\omega-\omega_{R})^{2}/(2\omega_{R})}(1+e^{-\beta\omega})$,
we calculate directly the interacting structure factor, $S_{\sigma\sigma^{\prime}}({\bf q},\omega)=zS_{\sigma\sigma^{\prime},1}^{\left(1\right)}+z^{2}[\Delta S_{\sigma\sigma^{\prime},2}+S_{\sigma\sigma^{\prime},2}^{\left(1\right)}]$,
once the fugacity $z$ is determined by the virial expansion for equation
of states.

\section{Results and discussions}

Considerable insight into the dynamic structure factor of a strongly
correlated Fermi gas can already be seen from Eq. (\ref{ddsf2}),
in which the spectrum is peaked roughly at $\omega_{R,m}=\omega_{R}/2$,
the recoil frequency for molecules. Therefore, the peak is related
to the response of molecules with mass $M=2m$. Eq. (\ref{ddsf2})
shows clearly how the molecular response develops with the modified
two-fermion energies and wave functions as the interaction strength
increases. In the BCS limit where $\Delta S_{\sigma\sigma^{\prime},2}$
is small, the response is determined by the non-interacting background
that peaks at $\omega_{R}$: see, for example, $S_{\sigma\sigma^{\prime},1}^{\left(1\right)}$
and $S_{\sigma\sigma^{\prime},2}^{\left(1\right)}$. In the extreme
BEC limit ($a\rightarrow0^{+}$), however, $\Delta S_{\sigma\sigma^{\prime},2}$
dominates. The sum in $\Delta S_{\sigma\sigma^{\prime},2}$ is exhausted
by the (lowest) tightly bound state $\phi_{rel}(x)\simeq\sqrt{2/a}e^{-x/a}$
with energy $\epsilon_{rel}\simeq-\epsilon_{B}\equiv-\hbar^{2}/(ma^{2})$.
The chemical potential of molecules is given by $\mu_{m}=2\mu+\epsilon_{B}$.
Therefore, the dynamic structure factor of fermions takes the form,
where $z_{m}=e^{\beta\mu_{m}}$ is the molecular fugacity of \begin{equation}
S_{\sigma\sigma^{\prime}}^{BEC}\simeq z_{m}B\sqrt{\frac{M}{\pi}}\exp\left[-\frac{\beta(\omega-\omega_{R,m})^{2}}{4\omega_{R,m}}\right].\label{dsfbec}\end{equation}
 This peaks at the molecular recoil energy $\omega_{R,m}$. As anticipated,
Eq. (\ref{dsfbec}) is exactly the leading virial expansion term in
the dynamic structure factor of non-interacting molecules (c.f. $S_{\sigma\sigma^{\prime},1}^{\left(1\right)}$).
It is clear that $S_{\uparrow\uparrow}({\bf q},\omega)\simeq S_{\uparrow\downarrow}({\bf q},\omega)$
in the BEC limit, since the spin structure in a single molecule can
no longer be resolved.

\begin{figure}
\begin{centering}
\includegraphics[clip,width=0.44\textwidth]{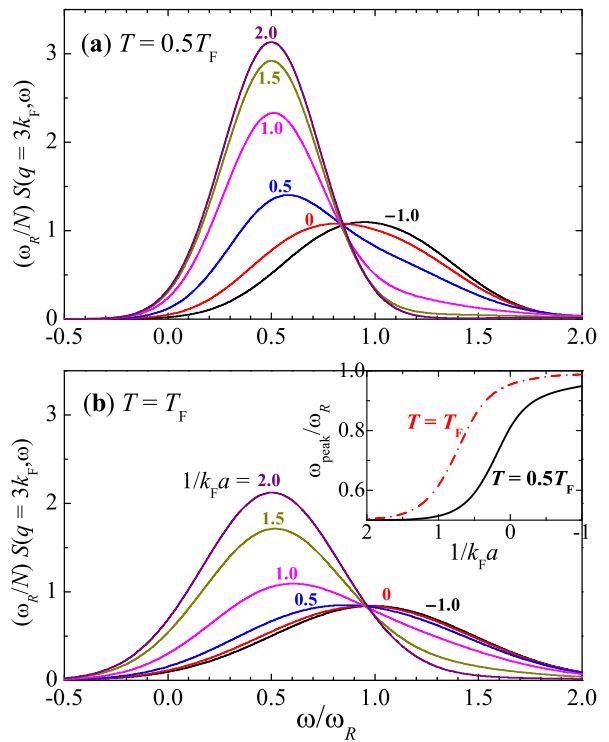} 
\par\end{centering}

\caption{(Color online). Evolution of the density structure factor with increasing
interaction strength $1/k_{F}a$ at $T=0.5T_{F}$ (a) and $T=T_{F}$
(b). The inset in (b) shows the peak position as a function of $1/k_{F}a$.}

\label{fig1} 
\end{figure}

To understand the intermediate regime, in Fig. 1 we report numerical
results for the total dynamic structure factor as the interaction
strength increases from the BCS to BEC regimes. In a trapped gas with
total number of fermions $N$, we use the zero temperature Thomas-Fermi
wave vector $k_{F}=(24N)^{1/6}/a_{ho}$ and temperature $T_{F}=(3N)^{1/3}\hbar\omega_{0}/k_{B}$
as characteristic units. In accord with the experiment \cite{swinexpt},
we take a large transferred momentum of $\hbar q=3\hbar k_{F}$. A
smooth transition from atomic to molecular responses is evident as
the interaction parameter $1/k_{F}a$ increases, in qualitative agreement
with the experimental observation (c.f. Fig. 2 in \cite{swinexpt}
). As shown in the inset, the transition shifts to the BEC side with
increasing temperature.

\begin{figure}
\begin{centering}
\includegraphics[clip,width=0.48\textwidth]{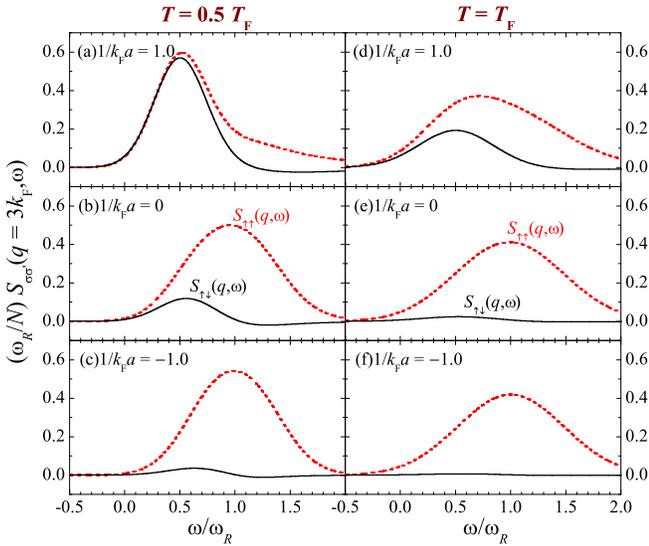} 
\par\end{centering}

\caption{(Color online). $S_{\uparrow\uparrow}({\bf q},\omega)$ (dashed lines)
and $S_{\uparrow\downarrow}({\bf q},\omega)$ (solid lines) at different
interaction strengths and at $T=0.5T_{F}$ (left column) and $T=T_{F}$
(right column).}

\label{fig2} 
\end{figure}

According to Eq. (\ref{ddsf2}), the molecular response in $\Delta S_{\uparrow\uparrow,2}$
and $\Delta S_{\uparrow\downarrow,2}$ should have the same order
of magnitude. However, the response is less obvious in the spin-parallel
channel because of the non-interacting background in $S_{\uparrow\uparrow}({\bf q},\omega)$.
As a result, $S_{\uparrow\downarrow}({\bf q},\omega)\simeq z^{2}\Delta S_{\uparrow\downarrow,2}$
provides an ideal probe for the formation of molecules. Experimentally,
$S_{\uparrow\downarrow}({\bf q},\omega)$ can be measured by a proper
choice of the detuning of the laser beams that results in different
couplings to the two spin components \cite{combescot}. In Fig. 2,
we show the evolution of $S_{\sigma\sigma^{\prime}}({\bf q},\omega)$
with increasing interaction strength (from bottom to top). At lower
temperatures (left column), $S_{\uparrow\downarrow}({\bf q},\omega)$
grows and becomes comparable to $S_{\uparrow\uparrow}({\bf q},\omega)$
at $1/k_{F}a=1.0$. It should be noted that our results at $T=0.5T_{F}$
are qualitatively reliable and are presented for illustrative purposes
only. We expect the predictions at $T=T_{F}$ to be more quantitative,
as estimated conservatively from the virial expansion of the equation
of states for a trapped strongly interacting Fermi gas \cite{unitaritycmp}.

\begin{figure}
\begin{centering}
\includegraphics[clip,width=0.44\textwidth]{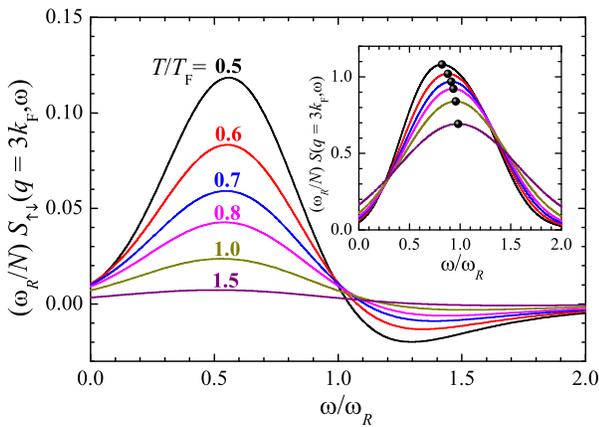} 
\par\end{centering}

\caption{(Color online). Temperature dependence of $S_{\uparrow\downarrow}({\bf q},\omega)$
at unitarity. The inset shows the red shift of the total dynamic structure
factor with decreasing temperature.}

\label{fig3} 
\end{figure}

Fig. 3 gives the temperature dependence of the DSF at unitarity. As
expected, $S_{\uparrow\downarrow}({\bf q},\omega)$ increases rapidly
with decreasing temperature. As a consequence, the peak of total dynamic
structure factor shifts towards the molecular recoil frequency, as
indicated in the inset. This red-shift of atomic peak was indeed observed
experimentally for a unitarity Fermi gas at $T\simeq0.4T_{F}$ \cite{private}.

\begin{figure}
\begin{centering}
\includegraphics[clip,width=0.48\textwidth]{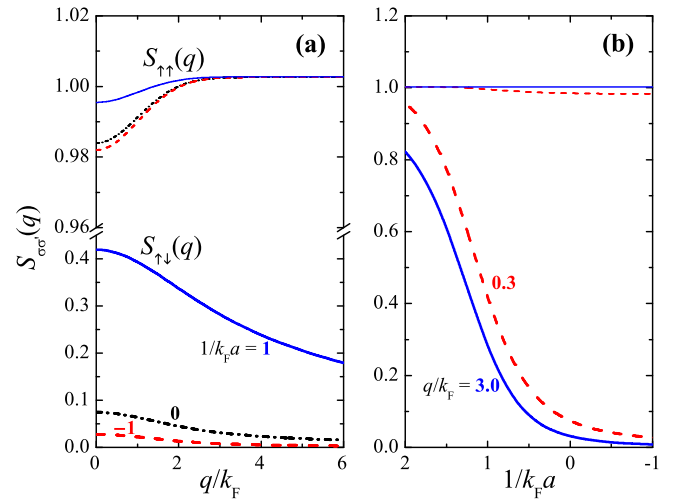} 
\par\end{centering}

\caption{(Color online). Dependence of static structure factors on the momentum
(a) and interaction strength (b) at $T=T_{F}$.}

\label{fig4} 
\end{figure}

We have so far restricted ourselves to cases with large $q$. The
momentum dependence can be conveniently illustrated by the \emph{static}
structure factor $S_{\sigma\sigma^{\prime}}({\bf q})=(2/N)\int d\omega S_{\sigma\sigma^{\prime}}({\bf q},\omega)$.
In Fig. 4, we present $S_{\sigma\sigma^{\prime}}({\bf q})$ as a function
of momentum (Fig. 4a) and interaction strength (Fig. 4b) at the Fermi
temperature $T_{F}$. The spin-antiparallel structure factor depends
strongly on momentum and interaction strength. In contrast, the spin-parallel
structure is nearly always unity, partly due to the autocorrelations
among identical spins. To better understand this, we consider the
short-range behaivor of the spin-parallel structure factor in real
space, $S_{\sigma\sigma}\left(\mathbf{{\normalcolor r}},\mathbf{r}'\right)\equiv(2/N)\left\langle \hat{n}_{\sigma}\left({\bf r}\right)\hat{n}_{\sigma}\left({\bf r}^{\prime}\right)\right\rangle $.
For parallel spins, there is no singularity as $\mathbf{r}'\rightarrow\mathbf{r}$.
Thus, to a good approximation $S_{\sigma\sigma}\left(\mathbf{{\normalcolor r}},\mathbf{r}'\right)\simeq(2/N)\left\langle \hat{n}_{\sigma}\left({\bf r}\right)\right\rangle \delta\left(\mathbf{r}-\mathbf{r}'\right)$,
where the delta function comes from the anticommutator of fermion
operators. After the Fourier transformation and average over the trap,
we obtain $S_{\sigma\sigma}({\bf q})\simeq1$.

\section{Conclusions}

In conclusion, we have developed a quantum virial expansion for dynamical
properties of strongly correlated systems and have computed the dynamic
density response of a strongly interacting Fermi gas at temperatures
$T\sim T_{F}$. The experimentally observed transition from atomic
to molecular response at low temperatures ($\sim0.1T_{F}$) has been
reproduced at a qualitative level. We anticipate that future Bragg
spectroscopy at high temperatures will lead to a quantitative agreement.
Alternatively, by including higher-order virial expansion functions,
we may extend the validity of our results closer to the characteristic
critical temperature $\sim0.2T_{F}$ \cite{unitaritycmp}. We emphasize
that our virial theory is efficient for investigating other basic
dynamical properties, such as the spectral function of single-particle
Green function. In this respect, it may shed light on solving the
paradox of the pseudogap phenomenon at unitarity \cite{akw}. We leave
this possibility in a future publication.

\section*{Acknowledgments}

We thank P. Hannaford and C. J. Vale for fruitful discussions. This
work was supported in part by the Australian Research Council (ARC)
Centre of Excellence for Quantum-Atom Optics, ARC Discovery Project
Nos. DP0984522 and DP0984637, NSFC Grant No. NSFC-10774190, and NFRPC
Grant Nos. 2006CB921404 and 2006CB921306.

\end{document}